\documentclass[twocolumn,superscriptaddress,showpacs]{revtex4}
\usepackage{graphicx}
\usepackage{epstopdf}
\usepackage{times}
\usepackage{bm}

\begin{document}
\title{Coherent control of dressed matter waves}
\author{Alessandro Zenesini}
\affiliation{Dipartimento di Fisica `E. Fermi', Universit\`a di
Pisa,} \affiliation{CNISM UdR Universit\`a di Pisa,}
\author{Hans Lignier}
\affiliation{Dipartimento di Fisica `E. Fermi', Universit\`a di
Pisa,}\affiliation{CNR-INFM, Dipartimento di Fisica `E. Fermi',
Largo Pontecorvo 3, 56127 Pisa, Italy}
\author{Donatella Ciampini}
\affiliation{Dipartimento di Fisica `E. Fermi', Universit\`a di
Pisa,} \affiliation{CNISM UdR Universit\`a di Pisa,}
\author{Oliver Morsch}
\affiliation{Dipartimento di Fisica `E. Fermi', Universit\`a di
Pisa,}\affiliation{CNR-INFM, Dipartimento di Fisica `E. Fermi',
Largo Pontecorvo 3, 56127 Pisa, Italy}
\author{Ennio Arimondo}\affiliation{Dipartimento di Fisica `E. Fermi', Universit\`a di
Pisa,} \affiliation{CNISM UdR Universit\`a di Pisa,}
\affiliation{CNR-INFM, Dipartimento di Fisica `E. Fermi', Largo
Pontecorvo 3, 56127 Pisa, Italy} 
\maketitle

{\bf By moving the pivot of a pendulum rapidly up and down one can
create a stable position with the pendulum's bob above the pivot
rather than below it~\cite{butikov_2001}. This surprising and
counterintuitive phenomenon is a widespread feature of driven
systems and carries over into the quantum world. Even when the
static properties of a quantum system are known, its response to
an explicitly time-dependent variation of its parameters may be
highly nontrivial, and qualitatively new states can appear that
were absent in the original system. In quantum mechanics the
archetype for this kind of behaviour is an atom in a radiation
field, which exhibits a number of fundamental phenomena such as
the modification of its $g$-factor in a radio-frequency
field~\cite{haroche_1970} and the dipole force acting on an atom
moving in a spatially varying light field~\cite{dalibard_1985}.
These effects can be successfully described in the so-called
dressed atom picture~\cite{cohen_1968}. Here we show that the
concept of dressing can also be applied to macroscopic matter
waves~\cite{eckardt_2008}, and that the quantum states of "dressed
matter waves" can be coherently controlled. In our experiments we
use Bose-Einstein condensates in driven optical lattices and
demonstrate that the many-body state of this system can be
adiabatically and reversibly changed between a superfluid and a
Mott insulating state~\cite{fisher_1989,jaksch_1998,greiner} by
varying the amplitude of the driving. Our setup represents a
versatile testing ground for driven quantum systems, and our
results indicate the direction towards new quantum control schemes
for matter waves.}

An atom in a radiation field can be described in the dressed atom
picture~\cite{cohen_1968} (or in equivalent approaches using,
e.g., Floquet quasienergy states) in which the modified properties
of the driven system arise from "dressing" the atom's electronic
states with the photons of the radiation field. This concept can
also be applied to macroscopic matter waves in driven periodic
potentials~\cite{eckardt_2008}, where the "dressing" is provided
by the oscillatory motion of the lattice potential. In analogy to
the dressed atom picture, such "dressed matter waves" can exhibit
new properties absent in the original system and thus allow
enhanced control of its quantum states. Here we demonstrate that
matter waves can be adiabatically transferred into a well-defined
Floquet quasienergy state of a driven periodic potential while
preserving their quantum coherence.

\begin{figure}[htp]
\includegraphics[width=8cm]{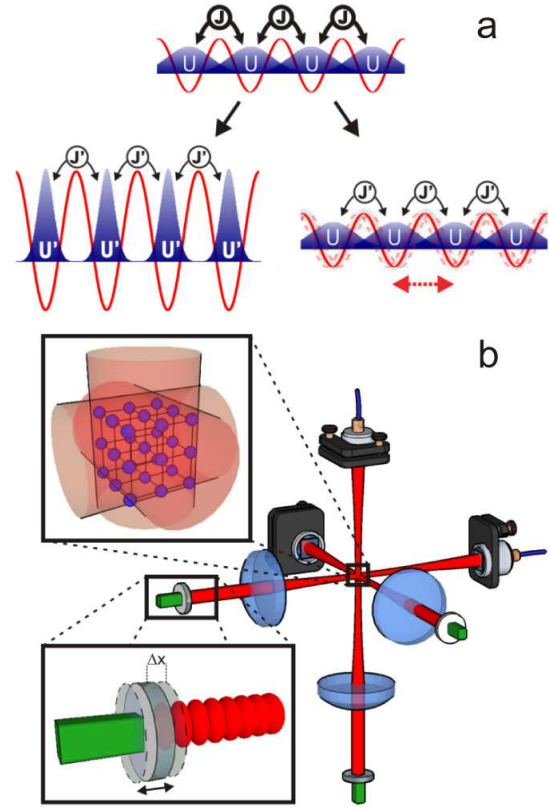}
\caption{{\bf a}, Principle of the coherent control of the
tunneling parameter $J$. A condensate in a lattice is
characterized by the tunneling parameter $J$ and the on-site
interaction $U$ (above). While changing the lattice depth results
in a variation of both $U$ and $J$ (left) to $U'$ and $J'$, strong
driving selectively changes $J$ to $J'=J_\mathrm{eff}$ (right).
{\bf b}, Experimental setup for a three-dimensional driven optical
lattice. The periodic potentials in the three spatial directions
are produced by retro-reflecting a focused laser beam off a mirror
mounted on a piezo-electric actuator.}
\end{figure}

Cold atoms in optical lattices~\cite{jaksch_1998} can be described
in the Bose-Hubbard model by the parameter $U/J$, where $J$ is the
hopping term relating to tunneling between adjacent sites, and $U$
is the on-site interaction energy (see Fig. 1a). When $U/J$ is
small, tunneling dominates and the atoms are delocalized over the
lattice, whereas a large value means that the interaction term is
large compared to $J$ and phase coherence is lost through the
formation of number-squeezed states with increased quantum phase
fluctuations. At a critical value of $U/J$ the system undergoes a
quantum phase transition to a Mott insulator state. Using optical
lattices one can tune $U/J$ by changing the lattice
depth~\cite{greiner,stoferle_2004}, which affects both $U$ and $J$
through the width of the on-site wavefunction (Fig. 1a).
Alternatively, by periodically shaking the lattice $J$ can be
suppressed~\cite{dunlap_86,holthaus_1992}. This principle, related
to the coherent destruction of tunneling in double-well
systems~\cite{grossmann_91,grifoni_1998}, was recently
experimentally demonstrated~\cite{lignier_2007,kierig_2008}. In
the driven system $J$ is replaced by an effective $J_\mathrm{eff}=
{\cal J}_0(K_0)J$, with ${\cal J}_0$ the zeroth-order Bessel
function and $K_0$ the driving strength (defined below). This
suggests that it should be possible to use $J_\mathrm{eff}$ in the
many-body Hamiltonian describing a BEC in a lattice and hence to
define an effective parameter
$U/J_\mathrm{eff}$~\cite{eckardt_2005}. In the following, we show
that this assumption is borne out by experiment.

In our experiments we created Bose-Einstein condensates of
$6\times 10^4$ atoms of $^{87}\rm Rb$, which were then
adiabatically loaded into the lowest energy band of an optical
lattice~\cite{gericke_2007}. The 1, 2 or 3-dimensional lattices
were realized by focusing linearly polarized laser beams ($\lambda
= 842\,\rm nm$) onto the BEC. Each lattice beam was
retro-reflected by a combination of a lens and a mirror (see
Fig.~1b), resulting in periodic potentials $V(x_i)=V_0\sin^2(\pi
x_i/d_L)$ along the three spatial directions, where
$d_L=\lambda/2$ is the lattice constant, $V_0$ the lattice depth
and $x_i=x,y,z$. The mirrors were mounted on piezo-electric
actuators that allowed us to sinusoidally shake each optical
lattice back and forth~\cite{ivanov_2008} with frequency $\omega$
(up to several kHz) and amplitude $\Delta x_i$. We define a
dimensionless driving strength $K_0=K/\hbar\omega=(\pi^2/2)(\Delta
x_i/d_L)(\omega/\omega_\mathrm{rec})$, where
$\omega_\mathrm{rec}=\hbar\pi^2/2md_L^2=2\pi\times
3.24\,\mathrm{kHz}$ is the recoil frequency (with $m$ the mass of
the $^{87}\mathrm{Rb}$ atoms).

\begin{figure}[htp]
\includegraphics[width=8cm]{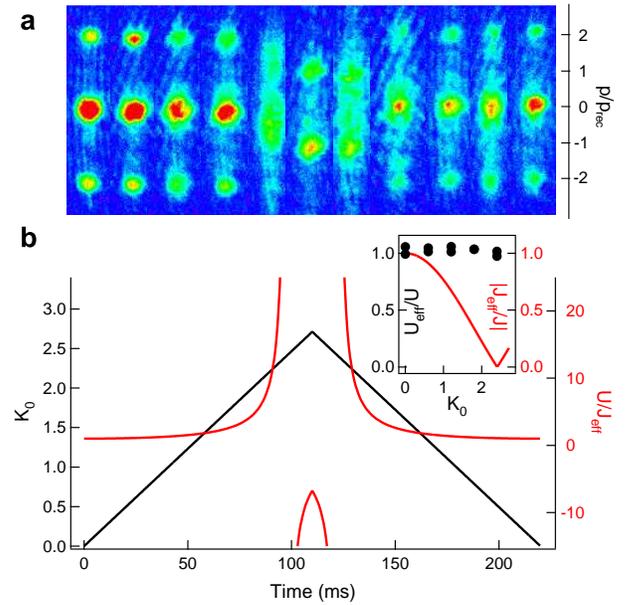}
\caption{Coherent control of a BEC inside a driven 1D lattice.
{\bf a}, The interference pattern after a time-of-flight reveals
that in the regions of the driving parameter $K_0$ for which
$|U/J_{\rm eff}|\ll 1$ phase coherence is preserved and a clean
three-peak structure is observed, whereas for $|U/J_{\rm eff}|\gg
1$ coherence is lost due to the formation of a number-squeezed
state. When $U/J_{\rm eff}<0$, a two-peaked structure at $\pm
p_\mathrm{rec}$ appears, indicating a negative value of
$J_\mathrm{eff}$ (the length scale of the interference patterns
has been converted into momentum units, with $1p_\mathrm{rec}$
corresponding to $138\,\mathrm{\mu m}$). The driving frequency was
$\omega/2\pi = 6\,\rm kHz$ and the lattice depth
$V_0=18\,E_\mathrm{rec}$.  {\bf b}, Time dependence of the driving
strength $K_0$ and $U/J_\mathrm{eff}$. {\it Inset:} The measured
(normalized) effective interaction parameter $U_\mathrm{eff}/U$ as
a function of $K_0$. For comparison, the (theoretical) behaviour
of the modulus of the effective tunneling parameter
$|J_\mathrm{eff}/J|$ is also shown.}
\end{figure}

In order to show that a BEC in a driven lattice (a) maintains its
phase coherence and (b) adiabatically follows changes in the
$K_0$, we performed preliminary experiments in one-dimensional
lattices (see Fig.~2). After loading a BEC into the lattice
($V_0=18\,E_\mathrm{rec}$, where
$E_\mathrm{rec}=\hbar\omega_\mathrm{rec}$ is the recoil energy),
$K_0$ was linearly increased from $0$ to $K_0=2.7$ in $113\,\rm
ms$ and back to $0$ in the same time. At times $t$ (where
$t=N\times 2\pi/\omega$ was an integer multiple of the driving
period) the lattice and the dipole trap were suddenly switched off
and the atoms were imaged on a CCD camera after $23.8\,\rm ms$ of
free fall. The interference pattern created by atoms originating
from different lattice wells (in our experiments around $40$ sites
were occupied) consisted of well-defined peaks when the condensate
was phase coherent over the entire lattice, whereas when phase
coherence was lost a broader, featureless pattern was observed.
Fig.~2a shows that when $J_{\rm eff}$ is large and hence $U/J_{\rm
eff}\ll 1$, the phase coherence persists for several tens of
milliseconds in spite of the strong driving. The appearance of a
stable, well-defined interference pattern proves that the BEC
occupies a single Floquet state of the driven system and
adiabatically follows that state as $K_0$ is varied. We also
verified that while $J_{\mathrm{eff}}$ changes with $K_0$, the
effective interaction parameter $U_\mathrm{eff}$ (inferred from
the the relative height of the side-peaks in the interference
pattern) remains constant (see the inset of Fig. 2b).

While for $K_0<2.4$ (for $K_0\approx2.4$, ${\cal J}_0(K_0)\approx
0$) the condensate occupies a Floquet state with quasimomentum
$q=0$ at the center of the Brillouin zone as reflected by an
interference pattern with a dominant peak at zero momentum and
sidepeaks at $\pm 2p_\mathrm{rec}=\pm 2\times h/\lambda$, for
$K_0>2.4$ (where $J_{\rm eff}$ is negative) two peaks at $\pm
p_\mathrm{rec}$ appear. This indicates that the Floquet state of
lowest mean energy now corresponds to $q=\pm p_\mathrm{rec}$ at
the edge of the Brillouin zone. Finally, when $K_0=2.4$ and hence
$J_{\rm eff}\approx 0$, $U/J_{\rm eff}\gg 1$ and phase coherence
is lost due to increased quantum phase fluctuations ({\it 22}).
When $K_0$ is reduced back to $0$ at the end of the cycle, the
initial interference pattern is restored almost perfectly,
suggesting that the response of the system to the parameter
variation was adiabatic.

\begin{figure}[htp]
\includegraphics[width=7cm]{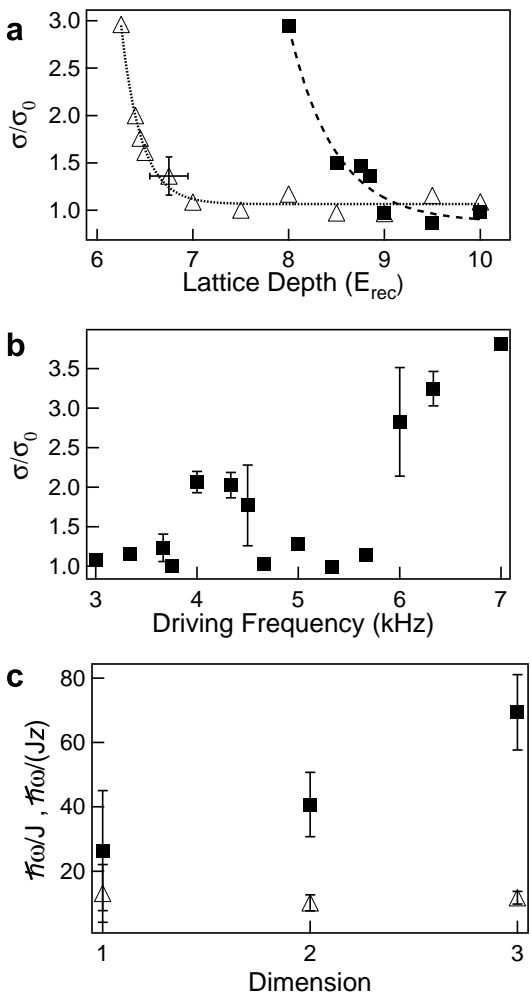}
\caption{Adiabaticity in driven optical lattices. {\bf a},
Dependence of the adiabaticity parameter $\sigma/\sigma_0$ (see
text) on the lattice depth in a one-dimensional lattice. The
adiabatic cycle consisted of a linear ramp $K_0=0$ to $K_0=2.3$ in
$15\,\mathrm{ms}$, a holding time at $K_0=2.3$ of
$200\,\mathrm{ms}$ and an identical ramp back to $K_0=0$. The
driving frequency was $3\,\mathrm{kHz}$ for the open triangles and
$6\,\mathrm{kHz}$ for the solid squares. {\bf b}, Dependence of
the adiabaticity on the driving frequency for a fixed lattice
depth $V_0=5.75\,E_\mathrm{rec}$. Here, the two linear ramps were
performed in $10\,\mathrm{ms}$ with a holding time of
$2\,\mathrm{ms}$. {\bf c}, Minimum value of $\hbar\omega/J$ (solid
squares) and $\hbar\omega/Jz$ (open triangles) for different
lattice dimensions. While $\hbar\omega/J$ increases with
increasing dimension $d$, the scaled parameter $\hbar\omega/Jz$
(with $z=2d$ the number of nearest neighbours) stays approximately
constant.}
\end{figure}

Since adiabaticity is a key concept in physics, we studied the
conditions for adiabaticity in our system more systematically.
This is important as the intuitive idea of an arbitrarily slow
change in one of the system's parameters allowing it to adjust its
state to the instantaneous parameter values at all times is no
longer valid in driven systems~\cite{eckardt_2008,hone_1997}. The
degree of adiabaticity in our experiments was measured by
performing cycles with triangular ramps for various lattice
depths, driving frequencies, ramp durations and dimensionalities
of the lattice. In order to compare the results for different sets
of parameters, at the end of the cycle we ramped down $V_0$ to
$4\,E_{rec}$ and measured the ratio of the width $\sigma$ of the
interference peak at $p=0$ and its width $\sigma_0$ for $K_0=0$,
which reflects any increase in energy and / or loss of coherence
during the cycle. The main results are summarized in Fig.~3.
Clearly, for fixed driving frequencies and ramp durations there
exist minimum lattice depths below which no adiabatic ramping is
possible (Fig. 3a), as indicated by the sharp increase in
$\sigma/\sigma_0$ below those values. This minimum is well-defined
and narrow and suggests a transition to a chaotic regime or
interband transitions induced by the driving. We also found that
for a given lattice depth the degree of adiabaticity depends
sensitively on the driving frequency (Fig. 3b). Again, interband
transitions may be responsible for the breakdown of adiabaticity
at frequencies above $6\,\mathrm{kHz}$, while other features such
as the partial breakdown between $4\,\mathrm{kHz}$ and
$4.5\,\mathrm{kHz}$ cannot be explained in this way. We also
investigated the dependence of the degree of adiabaticity on the
ramp duration keeping $V_0$ and $\omega$ constant. We found that
there exists an optimum ramp time of around $20\,\mathrm{ms}$
(depending slightly on $V_0$ and $\omega$).

Furthermore, we performed adiabaticity tests with two- and
three-dimensional lattices (see Fig. 3c) for which the BEC was
loaded into a 2D or 3D lattice as described above for the 1D case,
and the driving strength of the lattices was then ramped up and
down (using the same frequencies, phases and driving strengths for
all the lattices). Again, adiabatic ramps were possible for
certain sets of parameters. In particular, the minimum depth for
adiabaticity increased with increasing dimension $d$, whereas the
optimum ramp time decreased to a few milliseconds. Defining a
dimensionless parameter $\hbar\omega/J$, we found that for a given
$\omega$ the minimum value of the normalized ratio $\hbar
\omega/Jz$ (where $z=2d$ is the number of nearest neighbours) is
constant at about $12$ (Fig. 3c shows the mean values for
different driving frequencies and ramp times). While this suggests
that the ratio $\hbar \omega/zJ$ is useful for describing the
borderline in $J$ below which adiabatic control is possible at
constant $\omega$, we also found that changing $\omega$ and $J$
independently for a given $d$ does not always give the same value
(as indicated by the error bars in Fig. 3c). As already seen in
Fig. 3b, the conditions for adiabatic following depend on $J$ and
$\omega$ in a more complicated way. While these results are only a
first step towards understanding adiabatic following of Floquet
states, and more theoretical and experimental work needs to be
done, they nevertheless show that there are large regions in
parameter space for which adiabatic control is possible.

\begin{figure}[htp]
\includegraphics[width=9cm]{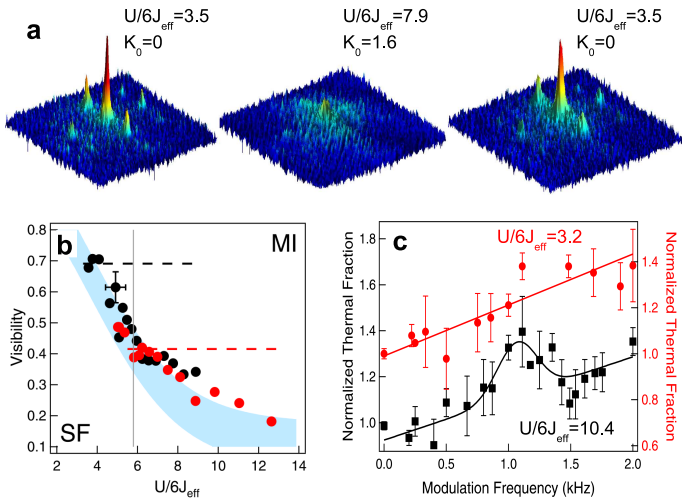}
\caption{Driving-induced Mott insulator transition in a 3D optical
lattice. {\bf a}, In a driven lattice of constant depth
($V=11\,E_{\mathrm{rec}}$, $\omega/2\pi=6\,\mathrm{kHz}$), $K_0$
was ramped from $0$ to $K_0=1.6$ in $4\,\mathrm{ms}$ and back
again. The disappearance of the interference peaks marks the onset
of the Mott insulator transition. {\bf b}, Visibility of the
interference pattern as a function of $U/6J_\mathrm{eff}$. For two
constant lattice depths $V=11\,E_{\mathrm{rec}}$ (black symbols)
and $V=12.2\,E_{\mathrm{rec}}$ (red symbols) , $K_0$ was varied.
The dotted black and red lines indicate the mean visibility after
returning to $K_0=0$ for the respective lattice depths. The blue
region is a fit (its width indicating the statistical error) to
the experimental data taken for a varying lattice depth without
driving, and the vertical grey line indicates the border between
the superfluid (SF) and Mott insulator (MI) regions. {\bf c},
Excitation spectrum measured by modulating $J_\mathrm{eff}$.
Plotted here is the thermal fraction (normalized to the thermal
fraction without excitation) calculated from a bimodal fit to the
condensate after ramping down the lattice to
$V_0=4\,E_\mathrm{rec}$. For the red circles $V_0=10.3$ with
$K_0=0.7$, while for the black squares $V_0=11$ with $K_0=1.8$.
The effective tunneling rate $J_\mathrm{eff}$ was modulated by
sinusoidally varying $K_0$ with a modulation depth of 0.8 and 0.3,
respectively. The solid lines are fits to guide the eye, and the
vertical scales have been offset for clarity.}
\end{figure}

We now turn to the driving-induced superfluid-Mott insulator
transition effected through an adiabatic variation of
$K_0$~\cite{eckardt_2005,creffield_2006}. We first loaded a BEC
into a 3D lattice of depth $V_0=11\,E_\mathrm{rec}$ and then
linearly ramped $K_0$ from $0$ to $K_0=1.6$. While in an undriven
lattice at $11\,E_\mathrm{rec}$ the BEC is in the superfluid
regime with $U/6J\approx 3.5$, the effective Bose-Hubbard
parameter $U/6J_\mathrm{eff}$ for the driven lattice at $K_0=1.6$
is around $7.9$, which is larger than the critical
value~\cite{jaksch_1998} of $5.4$ and hence the system is in the
Mott insulating phase~(see Fig.~4a). In this region, we observe a
distinct loss of phase coherence in the interference
pattern~\cite{greiner}. When $K_0$ is ramped back to $0$, the
interference pattern re-appears, proving that the transition was
induced adiabatically and that the BEC was not excited by the
driving. In order to have a more quantitative indication of the
phase transition, we have induced the Mott insulator transition in
two different ways: (a) by increasing $V_0$ in an undriven lattice
as in~\cite{greiner} and (b) by varying $K_0$ for a fixed lattice
depth. In Fig.~4b one clearly sees that the visibility of the
interference pattern~\cite{gerbier_2005} vanishes as
$U/6J_\mathrm{eff}$ is increased (and returns to its original
value after ramping $K_0$ back to $0$, as indicated by the
horizontal dashed lines). The dependence of the visibility on
$U/6J_\mathrm{eff}$ is the same for methods (a) and (b), strongly
indicating that in both cases the same many-body state is reached.
The independent control over $J_\mathrm{eff}$ also allowed us to
measure the excitation spectrum of the
system~\cite{stoferle_2004,clark_2006} by sinusoidally modulating
$J_\mathrm{eff}$ (rather than by modulating $V_0$, which also
changes $U$). While in the superfluid regime a gapless excitation
strength as a function of the modulation frequency appears, in the
Mott insulator regime we find a gapped spectrum (Fig. 4c).

Our results confirm and extend the role of cold atoms in optical
lattices as versatile quantum
simulators~\cite{greiner_2008,bloch_2008} and open new avenues for
the quantum control of cold atoms, thus establishing a link to
coherent control in other systems such as molecules in laser
fields~\cite{yamanouchi_2006} and Cooper pairs in Josephson
qubits~\cite{sillanpaa_2006}. The principles demonstrated here can
be straightforwardly extended to more than one driving
frequency~\cite{klumpp_2007} and to more complicated lattice
geometries such as superlattices~\cite{creffield_2007}.

\section*{Methods}
\subsection*{Driven optical lattices}
The driven or "spatially shaken" lattices (see Fig. 1a) were
realized by mounting the retro-reflecting mirror for each lattice
on a piezo-electric actuator (Queensgate Instruments, model
MTP15). These actuators were powered by three phase-locked
Stanford function generators producing a sinusoidal signal, the
amplitude of which could be controlled between $0$ and $10$ volts.
The response of the mirror-actuator couples had been previously
checked in an interferometric setup for the range of driving
frequencies used in the experiments (between $3\,\mathrm{kHz}$ and
$7\,\mathrm{kHz}$). Furthermore, the actuators could be calibrated
{\it in situ} using two different methods: \begin{enumerate} \item
By observing the interference pattern of a condensate released
from a driven one-dimensional lattice after a few milliseconds. We
repeated this experiment for increasing values of the driving
amplitude until the interference pattern was completely dephased.
This amplitude then corresponded to the point where
$J_\mathrm{eff}=0$ and hence $K_0=2.4$. Previously we had checked
that the spatial amplitudes of the oscillations of the
actuator-mirror couples were linear in the driving voltage (as
measured at the connections of the actuators), so having
calibrated the voltage $V$ for which $K_0(V)=2.4$ we could
extrapolate to the other values.
\item By observing the free expansion of a condensate in a driven
lattice~\cite{lignier_2007}. The condensate was allowed to freely
expand in the lattice direction by switching off one of the dipole
traps, and the width of the condensate was observed {\it in situ}
after a fixed expansion time. In this way, the Bessel-function
renormalization of the tunneling parameter $J_\mathrm{eff}={\cal
J}_0(K_0) J$ could be directly measured.
\end{enumerate}

\section*{Acknowledgments}Financial support by the E.U.-STREP "OLAQUI" and by a CNISM "Progetto Innesco 2007" is gratefully acknowledged. We thank J. Radogostowicz , C.
Sias and Y. Singh for assistance, and M. Holthaus and T. Esslinger
for discussions and a careful reading of the manuscript.
\section*{Competing financial interests}
The authors declare that they have no competing financial
interests.
\bibliographystyle{apsrmp}

\end{document}